\documentstyle[11pt]{article}

\newcommand{\half}{{\scriptstyle{\frac{1}{2}}}}
\newcommand{\msbar}{\overline{\rm MS}}
\begin{document}
\begin{titlepage}
\begin{flushright}
\hspace*{8.5cm} {\small DE-FG05-92ER40717-12}
\end{flushright}
\vspace*{18mm}
\begin{center}
               {\LARGE\bf The $16\half-N_f$ Expansion \\
                  and the Infrared Fixed Point \\
\vspace*{4mm}
               in Perturbative QCD}
\vspace{20mm}\\
{\Large P. M. Stevenson}
\vspace{18mm}\\
{\large\it
T.W. Bonner Laboratory, Physics Department,\\
Rice University, Houston, TX 77251, USA}
\vspace{25mm}\\
{\bf Abstract:}
\end{center}

    In QCD with $16\half-\epsilon$ massless quark flavours there
is an infrared fixed point with $\alpha_s/\pi = \frac{8}{321}\epsilon$,
in the limit $\epsilon \to 0_+$.  I develop the idea of Banks and Zaks
to expand about $N_f=16\half$.  This expansion is certainly useful for
$N_f=16, 15, 14, \ldots$, and arguably it can reach the
phenomenologically interesting case $N_f=2$, where it suggests that
$\alpha_s/\pi$ ``freezes'' to a value of order of magnitude $0.4$ in
the infrared.

\end{titlepage}

\setcounter{page}{1}

   {\bf 1.}  In QCD with $N_f$ massless flavours asymptotic freedom is
lost if $N_f$ exceeds $16\half$, which is where the first coefficient
of the $\beta$ function vanishes.  In my notation $\beta$ has the form:
\begin{equation}
\label{beta}
\beta(a) \equiv \mu \frac{da}{d\mu} = - ba^2(1+ca+c_2a^2+\ldots),
\end{equation}
where $a \equiv \alpha_s/\pi$, $b = (33-2N_f)/6$, and
$bc=(153-19N_f)/12$.  Since $b$ has been factored out, $c,c_2,\ldots$
will naturally each have a simple pole at $N_f=16\half$.  For $N_f$
just below $16\half$ there is a zero of the $\beta$
function at $a^* \approx -1/c$, a small positive value.  If the couplant
$a(\mu)$ lies between 0 and $a^*$ at some energy scale $\mu$, then it is
trapped in this range at all energies.  One then has weak coupling at all
energies \cite{cas,bz}.

     In 1982 Banks and Zaks \cite{bz} suggested an expansion in powers
of $16\half-N_f$.  The idea has since been used by White \cite{white}
to study the Pomeron in QCD.  Grunberg \cite{grun} (though in a context
I deplore) has discovered some important features of the expansion.
I argue here that this expansion is relevant to the real world with just
two light flavours.
It implies that, as a {\it perturbative} effect, the QCD couplant
``freezes'' in the infrared.  This corroborates Ref. \cite{us}, which
studied $R_{e^+e^-}$ in third-order ``optimized'' perturbation theory
\cite{opt}.

     Readers who find this hard to swallow should nevertheless read on:
This expansion is fun and it certainly offers insight into theories
with $16, 15, 14, \ldots$ flavours, whether or not one believes it is
directly useful for $N_f=2$.

     The following analysis stays entirely within the realm of
perturbation theory, and I shall discuss nonperturbative effects only
briefly at the end.

\vspace*{2mm}
     {\bf 2.} The natural expansion parameter for the Banks-Zaks (BZ)
expansion is
\begin{equation}
a_0 \equiv \frac{8}{321}(16\half - N_f),
\end{equation}
which is the limiting form of $a^* \sim -1/c$ as $N_f \to 16\half$ from
below.  Because the coefficient $8/321$ is so small,
$a_0$ is tiny ($\sim 1/80$) for $N_f=16$ and remains of modest size
(0.36) even for $N_f=2$.

     It is convenient to re-write perturbative coefficients, eliminating
$N_f$ in favour of $a_0$ \cite{grun}.  The first two $\beta$-function
coefficients are:
\begin{equation}
b = \frac{107}{8} a_0, \quad\quad c = -\frac{1}{a_0} + c_{1,0},
\end{equation}
with $c_{1,0} = 19/4$.  The higher-order $\beta$-function coefficients
are renormalization-scheme (RS) dependent.  In any `regular' scheme
their $N_f$ dependence is such that one may write them as
\cite{kat,grun}:
\begin{equation}
c_2 = \frac{1}{a_0} \left( c_{2,-1} + c_{2,0}a_0 + c_{2,1}a_0^2
+ c_{2,2}a_0^3 \right) ,
\end{equation}
and so on.  (Note that a term $c_{i,j} a_0^p$ can be assigned a degree
$i+j-p$, and all terms in any formula have the same degree.)
In the $\msbar$ scheme one has \cite{c2}:
\begin{equation}
c_2(\msbar) = \frac{1}{a_0} \left( -\frac{37117}{10272} +
\frac{243}{32}a_0 + \frac{34775}{1536}a_0^2 \right) .
\end{equation}

     Consider some perturbatively calculable physical quantity in QCD.
The prototypical example is $R_{e^+e^-}(Q)=3 \sum q_i^2 (1 + {\cal R})$,
where
\begin{equation}
\label{r}
{\cal R} = a (1 + r_1 a + r_2 a^2 + \ldots).
\end{equation}
In any `regular' RS the coefficients $r_i$ are $i$th-order polynomials
in $N_f$, and hence in $a_0$:
\begin{eqnarray}
r_1 & = & r_{1,0} + r_{1,1} a_0, \\
r_2 & = & r_{2,0} + r_{2,1} a_0 + r_{2,2} a_0^2,
\end{eqnarray}
etc..  All these coefficients are RS dependent.  In the
$\msbar$ scheme for the $R_{e^+e^-}$ case one has \cite{r2}
\begin{eqnarray}
r_1(\msbar(\mu=Q)) & = & \frac{1}{12} +
\frac{107}{32}(11 - 8 \zeta(3))a_0, \\
r_2(\msbar(\mu =Q)) & = & \left[ - \frac{12521}{288} + 13 \zeta(3) \right]
+ {\cal O}(a_0).
\end{eqnarray}
It is noteworthy that $\zeta(3)$ does not appear in $r_{1,0}$ and
$\zeta(5)$ does not appear in $r_{2,0}$.  (I have ignored the $(\sum q)^2$
term in $r_2$; its $N_f$ dependence depends on the electric charges
assigned to the ficticious extra quarks.)

\vspace*{2mm}
     {\bf 3.} Consider first the BZ expansion for
${\cal R}^* \equiv {\cal R}(Q \to 0)$.  One first solves the fixed-point
condition $\beta(a^*) =0$ for $a^*$ as a series in $a_0$.  One then
substitutes in Eq. (\ref{r}), again expanding in powers of $a_0$.
To leading order ${\cal R}^* = a^* = a_0$, while at second order one has
\cite{grun}
\begin{equation}
\label{astar}
a^* = a_0 [ 1 + (c_{2,-1} + c_{1,0}) a_0 + {\cal O}(a_0^2) ],
\end{equation}
and hence
\begin{equation}
\label{rst}
{\cal R}^* = a_0 [ 1 + (r_{1,0} + c_{2,-1} + c_{1,0}) a_0 +
{\cal O}(a_0^2) ].
\end{equation}
The sum of $r_{1,0}$ and $c_{2,-1}$ is RS invariant, as I show later.
($c_{1,0}=19/4$ is invariant.)  In the $e^+e^-$ case one obtains
${\cal R}^* = a_0 ( 1 + 1.22 a_0 + \ldots )$, so the correction is relatively
modest.  At $N_f=2$ the correction is about 44\%.  While one cannot give
too much credence to the quantitative result, the qualitative message of
leading order remains; there seems to be a fixed point of modest size.
Figure 1 shows ${\cal R}^*$ as a function of $N_f$ and compares first-
and second-order BZ results with the OPT results from Ref. \cite{us}.

     The crucial test of this interpretation will come at next order.
A straightforward calculation yields the coefficient of the next order
term in (\ref{rst}), which is
\begin{equation}
\label{coef3}
\left( c_{1,0} + 2 c_{2,-1} + 2 r_{1,0} \right)
\left( c_{1,0} + c_{2,-1} \right)
+ r_{1,1} + r_{2,0} + c_{2,0} + c_{3,-1}.
\end{equation}
In the $e^+e^-$ case this reduces to $-18.25 + c_{3,-1}(\msbar)$.
For the expansion to be credible one needs $c_{3,-1}(\msbar)$
to be in the range, say, $+13$ to $+21$.  I expect $c_{3,-1}$ to
be found in the lower end of this range, thereby reducing ${\cal R}^*$,
and bringing it into better agreement with the OPT results.  A calculation
of the 4th-order $\beta$ function coefficient would test this prediction.

    Note that $n$th order in the BZ expansion requires $n+1$ terms in
the $\beta$ function, but only $n$ terms in ${\cal R}$.  Thus, in terms
of diagrammatic information used, it is intermediate between $n$th
and $(n+1)$th order of ordinary perturbation theory.

    The coefficients in the BZ expansion of ${\cal R}^*$ are RS invariant.
One can prove this by considering the RS invariants $\rho_i$ \cite{opt}
and expanding them in powers of $a_0$.   From the leading $1/a_0$ term in
$\tilde{\rho_2} \equiv \rho_2 + c^2/4$ $\equiv r_2 + c_2 - r_1^2 - c r_1$
one sees that $r_{1,0} + c_{2,-1}$ is invariant, as claimed earlier.
{}From the subleading part of $\tilde{\rho_2}$ one finds that
$r_{2,0}+c_{2,0}-r_{1,0}^2+r_{1,1}-c_{1,0}r_{1,0}$ is invariant.
Then from the leading $1/a_0$ term in
$\rho_3 \equiv r_3+\half c_3 - r_1(c_2+3r_2-2r_1^2-\half c r_1)$ one
finds that $c_{3,-1}-2r_{1,0} c_{2,-1}-r_{1,0}^2$ is invariant.
It is then straightforward to show that the combination in Eq.
(\ref{coef3}) is RS invariant.  It also follows that
$c_{2,-1}^2+c_{3,-1}$ is invariant \cite{grun}.

\vspace*{2mm}
     {\bf 4.} Next, consider a BZ expansion for ${\cal R}$ at a
general $Q$.  There are three preliminary steps.  Firstly, one integrates
the $\beta$-function equation to obtain:
\begin{equation}
\label{intbeta}
b \ln(\mu/{\tilde{\Lambda}}) = \lim_{\delta \to 0} \left[ \,
\int_{\delta}^{a} \!\frac{d x}{ \hat{\beta}(x)} + {\cal C}(\delta) \right],
\end{equation}
where $\hat{\beta}(x) \equiv \beta(x)/b$, and $\tilde{\Lambda}$ is a
constant with dimensions of mass.  The constant of integration,
${\cal C}(\delta)$, must be suitably singular in the limit $\delta \to 0$,
and I choose \cite{opt}
\begin{eqnarray}
{\cal C(\delta)} & = &
\mbox{\boldmath $P$} \int_{\delta}^{\infty} \! \frac{d x}{x^2(1 + c x)},
\nonumber \\
\label{const}
& = & \frac{1}{\delta} + c \ln \delta + c \ln \mid \! c \! \mid +
{\cal O}(\delta),
\end{eqnarray}
where {\boldmath $P$} (principal value) is specified because of the pole
at $x=-1/c$ when $c$ is negative.  This choice amounts to a definition,
in a general RS, of the $\tilde{\Lambda}$ parameter.  The commonly used
$\Lambda$ parameter \cite{bbdm} is defined in a less natural way, and
is related by an RS-invariant, but $N_f$-dependent factor;
$\ln (\Lambda/\tilde{\Lambda}) = (c/b) \ln \,(2 \!\mid\! c \!\mid\! /b)$.
The two $\Lambda$'s become infinitely different in the limit
$N_f \to 16\half$.

     Secondly, recall that for each physical quantity ${\cal R}$ there is
an RS invariant \cite{opt}:
\begin{equation}
\rho_1 \equiv b \ln (\mu/\tilde{\Lambda}) - r_1.
\end{equation}
The renormalization scale $\mu$ cancels out because the coefficient
$r_1$ always contains a $b \ln(\mu/Q)$ piece.  Furthermore, the
$\tilde{\Lambda}$ parameter is scheme dependent in a way that exactly
cancels the scheme dependence of the remaining part of $r_1$ \cite{cg,opt}.
($\rho_1$ must be regarded {\it as a whole}; splitting it into
pieces spoils RS invariance.  It {\it cannot} be written as $A + Ba_0$
with $A$ and $B$ being RS invariant \cite{cs}.  The reason is the $1/b$
factor in the RS transformation of $\tilde{\Lambda}$:  For two schemes
related by $a^{\prime} = a(1 + v_1 a + \ldots)$ one has
$\ln(\tilde{\Lambda}^{\prime}/\tilde{\Lambda}) = v_1/b$ \cite{cg}.
For `regular' schemes $v_1$ is linear in $N_f$, but is otherwise
arbitrary.  The other $\rho_i$ invariants {\it can} be split into
different orders in $a_0$ because they are not $Q/\tilde{\Lambda}$
dependent.)  One may think of $\rho_1$ as
$b \ln Q/\tilde{\Lambda}_{{\rm eff}}$, where
$\tilde{\Lambda}_{{\rm eff}} \equiv \tilde{\Lambda} \exp(r_1/b)$
is a scale specific to the particular physical quantity ${\cal R}$.
Each $\tilde{\Lambda}_{{\rm eff}}$ can be related in an exactly known
way, once the corresponding $r_1$ has been calculated, to the universal
$\tilde{\Lambda}$ of some reference scheme (say, $\msbar$), which plays
the role of the single free parameter of the theory.

     Thirdly, let us reconnoitre the BZ limit.  Since $a$ is trapped between
$0$ and $a^*$, it is at most of order $a_0$.  The first two
terms in the $\beta$ function dominate, so Eq. (\ref{intbeta}) gives
\begin{equation}
\label{intbeta2}
b \ln (\mu/\tilde{\Lambda}) = \frac{1}{a} +
c \ln \left| \frac{ca}{1+ca} \right| + {\cal O}(c_2 a).
\end{equation}
Combining the two last equations yields $\rho_1$ in terms of $a$.
The dominant terms are of order $1/a_0$, so one may discard the
${\cal O}(c_2 a)$ and $r_1$ terms, which are of order unity.  Since
in the BZ limit $c \sim -1/a_0$, ${\cal R} \sim a$ and
${\cal R}^* \sim a^* \sim a_0$, the limiting form is:
\begin{equation}
\label{bzlim}
\rho_1 = \frac{1}{{\cal R}} +
\frac{1}{{\cal R}^*} \ln \left( \frac{{\cal R}^*-{\cal R}}{{\cal R}} \right).
\end{equation}
Inverting this equation (numerically) would give ${\cal R}$ as a function
of $\rho_1 = b \ln Q/\tilde{\Lambda}_{{\rm eff}}$, and hence as a function
of $Q$.  The resulting function ${\cal R}(Q)$ is RS invariant, and is
universal, except that $\tilde{\Lambda}_{{\rm eff}}$ depends on the specific
physical quantity considered.  ${\cal R}(Q)$ exhibits both asymptotic-freedom
as $Q \to \infty$ and ``freezing'' behaviour, ${\cal R}(Q) \to {\cal R}^*$,
as $Q \to 0$.

     The formulation of a BZ expansion for ${\cal R}$ at a general $Q$ is
not a completely unambiguous matter.  ${\cal R}(Q)$ is not expressible as
a simple power series in $a_0$, so some thought is required in deciding
how precisely to define the $n$th-order approximant.  The important point,
as with any approximation, is to reconcile and make best use of all
available information.  Simply integrating the $\beta$-function equation
and then expanding in powers of $a_0$ produces correction terms with
$(a_0 - {\cal R})$ denominators.  Higher orders bring in ever more
singular terms.  However, these terms simply arise from an expansion of
$\ln({\cal R}^* - {\cal R})$, reflecting the fact that the fixed point
${\cal R}^*$ does not stay at $a_0$, but is itself a series in $a_0$.
Therefore it is sensible to organize the expansion to reflect this.

     Thus, before performing the integration in (\ref{intbeta}), I first
re-write the $1/\hat{\beta}(x)$ integrand:
\begin{equation}
\frac{1}{- x^2(1 + c x + c_2 x^2 + \ldots)} =
\frac{-a^*}{x^2(a^*-x)P(x)},
\end{equation}
ensuring that the pole is in the right place.  Next, I express it in partial
fractions:
\begin{equation}
\label{pf}
\frac{1}{\hat{\beta}(x)} =
-\frac{1}{x^2} + \frac{c}{x} - \frac{1}{\hat{\gamma}^*(a^*-x)}
+ H(x).
\end{equation}
The coefficients of the first three terms are determined by the $x \to 0$
and $x \to a^*$ limits.  Hence, $\hat{\gamma}^*$ is $\gamma^*/b$, where
$\gamma^*$ is the slope of the $\beta$ function at the fixed point:
\begin{equation}
\gamma^* \equiv \left. \frac{d \beta(x)}{d x} \right|_{x=a^*}
= - b a^*(1 + 2 c a^* + 3 c_2 {a^*}^2 + \ldots ).
\end{equation}
The remainder term can be expanded as a power series,
$H(x) = H_0 + H_1 x + \ldots$.

     In $n$th order of the BZ expansion one may truncate the $\beta$
function after $n+1$ terms.  In that case $H(x)=Q(x)/P(x)$, where
$P(x)$ and $Q(x)$ are polynomials of degree $n-1$ and $n-2$, respectively.
(For $n=1$, $Q(x)$ vanishes.)  The coefficients of $P(x)$ are of order
unity as $a_0 \to 0$.  The coefficients of $Q(x)$ are of order $a_0$
because of cancellations that make both $c+1/\hat{\gamma}^*$ and
$a^*/\hat{\gamma}^*-1$ of order $a_0$.  Thus, $H(x)$ has coefficients
of order $a_0$.  [$H_0$, for instance, is
$ a_0(c_{4,-1}+2c_{2,-1}c_{3,-1}+c_{2,-1}^3) + {\cal O}(a_0^2)$.]
In $n$th order ($n \ge 4$) of the BZ expansion one needs coefficients
up to $H_{n-4}$: for the first three orders one can drop $H(x)$
altogether.

     It is now simple to perform the integration in (\ref{intbeta})
and use (\ref{const}) to obtain $\rho_1$ in terms of $a$ and $a^*$.
One may then eliminate $a$ and $a^*$ in favour of ${\cal R}$ and
${\cal R}^*$, working to the appropriate order.  This last step can
be short circuited by noting that the final result must be RS invariant,
and so, without loss of generality, one may choose to work in the RS
in which ${\cal R} = a$.  Thus, the result in $n$th order of the BZ
expansion can be expressed as:
\begin{equation}
\label{bzq}
\rho_1 = \frac{1}{\cal R} + \frac{1}{\hat{\gamma}^{*(n)}}
\ln \left( 1 - \frac{{\cal R}}{{\cal R}^{*(n)}} \right)
+ c \ln ( \mid\! c \! \mid {\cal R} ) +
\sum_{i=0}^{n-4}\frac{H_i {\cal R}^{i+1}}{(i+1)},
\end{equation}
where ${\cal R}^{*(n)}$ and $\hat{\gamma}^{*(n)}$ are the $n$th-order
approximations to ${\cal R}^*$ and $\hat{\gamma}^*$, respectively.
For small ${\cal R}$ (i.e., at large $Q$) this formula will agree with
$(n+1)$th-order perturbation theory to the appropriate order in ${\cal R}$
and $a_0$.  For $\hat{\gamma}^*$ a straightforward calculation gives
\begin{equation}
\label{gamma}
\hat{\gamma}^* = \frac{\gamma^*}{b} = a_0 (1+ c_{1,0} a_0 +
(c_{1,0}^2-c_{2,-1}^2-c_{3,-1}) a_0^2 + {\cal O}(a_0^3) ).
\end{equation}
$\gamma^*$ is the `critical exponent' that governs the manner in which
${\cal R}$ approaches ${\cal R}^*$ in the $Q \to 0$ limit \cite{gross};
${\cal R}^* - {\cal R} \sim {\rm const.} \, Q^{\gamma^*}$.
It should therefore be RS invariant.  Actually, this statement must be
qualified \cite{gross,chyla}; $\gamma^*$ is invariant within the sub-class
of RS's whose relation to the $a={\cal R}$ scheme is not singular at
the fixed point.  The coefficients in (\ref{gamma}) are the ``universal''
invariants (universal in that they are independent of the particular
quantity ${\cal R}$) discovered by Grunberg \cite{grun}.

     Results for ${\cal R}(Q)$ in first and second order are shown in
Figs. 2 and 3.  One could also obtain hypothetical third-order results
by guessing a value for $c_{3,-1}$.  For a value around 15, for
instance, the $Q=0$ values are similar to the first-order results, with
the shape of the curves being more like second order.

     The first-order results use (\ref{bzq}) which, unlike the earlier
form (\ref{bzlim}), retains the whole of $c$ in the
$c \ln ( \mid\! c \! \mid {\cal R} )$ term.  This seems sensible
because it avoids spoiling the behaviour at large $Q$, known
from ordinary perturbation theory.  One must know $c$ in order to
obtain even the leading-order BZ result, in any case.  Another point
is that (\ref{bzq}) puts the corrections to the coefficient of
$\ln (1-{\cal R}/{\cal R}^*)$ in the $\hat{\gamma}^*$ denominator,
rather than re-expanding them into the numerator.  This is sensible
because one knows that $\gamma^*$, the slope of $\beta(a)$ at $a^*$,
must be positive, as is obvious from a sketch of the $\beta$ function.

     For $N_f=16,15,14,\ldots$ the theory becomes ``almost scale
invariant.''  ${\cal R}$ remains constant over a huge range of $Q$ about
$\tilde{\Lambda}_{{\rm eff}}$.  This is because
$\rho_1= (107/8) a_0 \ln(Q/\tilde{\Lambda}_{{\rm eff}})$ remains close
to zero.  Only when $Q/\tilde{\Lambda}_{{\rm eff}}$
becomes extremely small does ${\cal R}(Q)$ abruptly rise up to
${\cal R}^*$, while only when $Q/\tilde{\Lambda}_{{\rm eff}}$ becomes
very large does ${\cal R}(Q)$ slowly decrease, as required by
asymptotic freedom.  The constant value around
$\tilde{\Lambda}_{{\rm eff}}$ is, in the BZ limit,
${\cal R}^*/(1+\chi) \sim 0.78 {\cal R}^*$,
where $\ln \chi + \chi + 1 = 0$.  The region over which ${\cal R}$
stays within 10\% of this value is roughly for
$Q/\tilde{\Lambda}_{{\rm eff}}$ between the two extremes
$\exp(\pm0.04/a_0^2)$.  Thus, this region is very extensive for
$a_0 < 0.1$ ($N_f \ge 13$) and is noticeable up to $a_0 \sim 0.2$
($N_f \sim 9$).  For smaller $N_f$ the region
$Q \sim \tilde{\Lambda}_{{\rm eff}}$ becomes, on the contrary, a
region of rapid variation of ${\cal R}(Q)$: asymptotic freedom sets
in quickly above $\tilde{\Lambda}_{{\rm eff}}$, while infrared
``freeze-out'' occurs just below $\tilde{\Lambda}_{{\rm eff}}$.
The cause of this different behaviour is the much larger value of the
critical exponent $\gamma^*$ at smaller $N_f$.

\vspace*{2mm}
{\bf 5.}  The above analysis has been entirely within a perturbative
framework.\footnote{
The BZ limit might also be a useful way to explore nonperturbative
effects, such as instantons, in a controlled weak-coupling context.
}
Its main message is that the couplant ``freezes'' at low
energies as a {\it perturbative} effect.  This property is manifest
for $N_f$ close to $16\half$ and the BZ expansion implies that it extends
to all lower values of $N_f$, albeit with diminishing accuracy.
Because of ``freezing'' one obtains finite perturbative predictions
at low energies, without resort to such notions as constituent-quark
masses, effective gluon masses, condensates, etc..   I do not claim that
these predictions are {\it right} --- nonperturbative effects certainly do
exist at low energies ---  but I do argue that these predictions are
{\it meaningful} and have some predictive power.

    Let me first remark that ``freezing'' is not incompatible with
confinement: there is no evidence that confinement necessarily requires
the couplant to become infinite in the infrared.  (Gribov's ideas
\cite{gribov}, for instance, explicitly involve a ``freezing'' of the
couplant.)  Confinement and chiral-symmetry-breaking phenomena are
associated with terms invisible to perturbation theory, such as
$\exp(-1/ba)$, whose Taylor expansion is $0+0+0+\ldots$.
Even in the $N_f=16$ theory, where the couplant is always very small,
perturbation theory is probably not the whole story.  Nonperturbative
effects in that case may be extremely tiny but they could still be
qualitatively decisive in providing confinement and chiral symmetry
breaking.  (However, if the confinement radius is enormous relative to
any feasible experiment, then the physicists of an $N_f=16$ world might
well regard confinement as an irrelevant notion.)

    In the real world there are certainly nonperturbative effects of
crucial importance at low energies.  The perturbative predictions
are smooth functions, whereas the data is characterized by a fine
structure of hadronic thresholds and resonances.  However, it is natural
to conjecture that data and perturbative theory can be compared if some
suitable `smearing' procedure is used to `average out' the nonperturbative
effects.  This is an old idea (e.g. \cite{pqw}) that has appeared in many
forms.  I am only proposing that the game can be extended right down to
zero energy.  Indeed, such a comparison with $R_{e^+e^-}$ data proves to
be very successful \cite{us}.  The hypothesis is worth pursuing because it
potentially enlarges our predictive power.  It also changes the nature of
the debate about how perturbative and nonperturbative effects fit together:
The question is no longer ``How must we modify perturbation theory to
prevent it giving nonsense for $Q \le \tilde{\Lambda}$?'' but instead
becomes ``How do nonperturbative effects modulate the smooth perturbative
result at low energies to produce fine structure?''

     Quark masses were ignored in the preceding analysis.  However, it
should be straightforward to include running masses in the usual way
\cite{gp}.  The $c,b,t$ quarks clearly decouple at energies below the
$\tilde{\Lambda}$ scale; the $s$ quark is a borderline case; but it seems
reasonable to treat the $u,d$ quarks as approximately massless:
Their running masses are of order 5--10 MeV at a renormalization scale of
1 GeV \cite{qmass}, and although the running masses increase at lower
scales this rise is tempered by the freezing of $a$.  In any case, the
BZ-expansion results are rather insensitive to whether the effective $N_f$
for $Q \le \tilde{\Lambda}$ is taken to be 3 or 2 or 0.  (It would not
be appropriate to invoke constituent quark masses, which have no
well-defined connection with the parameters of the underlying Lagrangian
\cite{qmass}.  They represent a phenomenological attempt to parametrize
some of the nonperturbative effects.)

     ``Freezing'' has long been a popular, {\it ad hoc} hypothesis, and
it has been invoked in a great variety of successful phenomenology.
The low-energy values for $\alpha_s/\pi$ so obtained lie,
with remarkable consistency, in the range $0.2$ to $0.3$.  (See Ref.
\cite{us} for a brief literature survey.)  Usually ``freezing''  is
blamed on unspecified ``non-perturbative effects.''  The BZ and OPT
results imply, however, that ``freezing'' is a phenomenon present in
perturbation theory at second or third order.  The BZ expansion implies
that ``freezing'' is not attributable to $\pi^2$ terms; these only appear
in the ${\cal O}(a_0^2)$ part of the $r_2$ coefficient.  It also implies
that ``freezing'' is universal, regardless of spacelike/timelike $Q^2$,
and should occur for any perturbatively expandable physical quantity.

\vspace*{2mm}
\begin{center}
{\bf Acknowledgements}
\end{center}

\vspace*{-1.5mm}
    I am grateful to Alan White for introducing me to the concept of
the $16\half-N_f$ expansion.  I thank Ji\v{r}\'{\i} Ch\'yla and
Andrei Kataev for correspondence, and Frank Close, Poul Damgaard,
Valery Khoze, Alan Mattingly, Chris Maxwell, and Graham Shore for
discussions.

This work was supported in part by the U.S. Department of Energy under
Grant No. DE-FG05-92ER40717.

\end{document}